\documentstyle[12pt]{article}

\global\arraycolsep=1pt
\newcommand{\beq}{\begin{equation}}
\newcommand{\eeq}{\end{equation}}
\newcommand{\beqa}{\begin{eqnarray}}
\newcommand{\eeqa}{\end{eqnarray}}
\newcommand{\CR}{\nonumber \\}

\newcommand{\del}{\partial}

\newcommand{\trace}{{\hbox{Tr}}}

\newcommand{\hi}{\hat{i}}
\newcommand{\hj}{\hat{j}}
\newcommand{\hk}{\hat{k}}
\newcommand{\e}{\epsilon}

\begin{document}

\begin{titlepage}
\begin{center}
{\Large \bf
 Octonionic Yang-Mills Instanton \\
on Quaternionic Line Bundle of Spin(7) Holonomy}
\lineskip .75em
\vskip2.0cm
{\large Hiroaki Kanno\footnote{e-mail: kanno@math.sci.hiroshima-u.ac.jp}}
\vskip 1.5em
{\large\it Department of Mathematics, Faculty of Science \\
Hiroshima University, Higashi-Hiroshima, 739-8526, Japan}
\vskip1cm
{\large Yukinori Yasui\footnote{e-mail: yasui@sci.osaka-cu.ac.jp}}
\vskip 1.5em
{\large\it Department of Physics, Osaka City University \\
Sumiyoshi-ku, Osaka 558-8585, Japan}
\end{center}
\vskip1.5cm
\begin{abstract}
The total space of the spinor bundle on the four dimensional sphere $S^4$
is a quaternionic line bundle that admits a metric of $Spin(7)$ holonomy.
We consider octonionic Yang-Mills instanton on this eight
dimensional gravitational instanton. This is a higher dimensional
generalization of (anti-)self-dual instanton on the Eguchi-Hanson space.
We propose an ansatz for $Spin(7)$ Yang-Mills field and derive a system
of non-linear ordinary differential equations. The solutions are
classified according to the asymptotic behavior at infinity.
We give a complete solution, when the gauge group is reduced to
a product of $SU(2)$ subalgebras in $Spin(7)$. The existence of more general
$Spin(7)$ valued solutions can be seen by making an asymptotic expansion.
\begin{flushleft}
\end{flushleft}
 
\end{abstract}
\end{titlepage}

\section{Introduction}

Instantons and soliton solutions have played a prominent role
in our understanding of non-perturbative dynamics and dualities
in gauge field theories and string theory.
It has been observed that fundamental examples of topological solutions
are associated with the four Hopf fibrations of spheres,
which in turn are related to the division algebras of
real numbers {\bf R}, complex numbers {\bf C}, quaternions {\bf H}
and octonions {\bf O} \cite{Trau}\cite{GKS}.
The kink solution in $1+1$ dimensions,
the Dirac monopole in three dimensions and the $SU(2)$ Yang-Mills
instanton in four dimensions correspond to the first three algebras.
In this paper we will bring into focus eight dimensional instantons
that corresponds to the octonions.

If the theory is promoted to a supersymmetric (SUSY) theory,
these topological solutions obtain a new feature.
That is, they are characterized as BPS states that preserve a fraction of SUSY.
Roughly speaking a first order soliton equation is
 a \lq\lq square root\rq\rq\ of the equation of motion for bosons
and hence appears in the SUSY transformations of fermions.
Solutions that make the SUSY variation of fermions vanishing
give purely bosonic configrations which preserve (or break)
some portions of supersymmetry.
It is interesting that there is also a relation between
the four division algebras and the existence of supersymmetric
pure Yang-Mills theory and superstring theory
in $d=3,4,6,10$ \cite{Ev}\cite{BST}.
Thus we see there are amusing links among
instanton, supersymmetry and the division algebra. From this
viewpoint octonionic instanton that will be featured in the following
is related to SUSY Yang-Mills and superstring theory in ten dimensions.

Topological quantum field theories enter naturally in these connections.
Especially, topological Yang-Mills theories in two and four dimensions are,
respectively, associated with the complex numbers and the quaternions.
Furthermore, one can construct an eight dimensional cohomological
Yang-Mills theory based the octonionic instanton
equation \cite{BKS}\cite{ALS}, though it makes sense
only on a manifold of restricted holonomy.
It is promising that this BRST cohomological theory probes
the moduli space of the octonionic instanton equation.
But how this is achieved actually depends on our finding
an appropriate compactification of the moduli space.
In the case of four dimensional instantons we need
the point-like (ideal) instantons to compactify the moduli space.
In \cite{HK} it has been argued that a natural higher
dimensional analogue of ideal (point-like) instanton is that lives
on the normal bundle over a supersymmetric
cycle (or a calibrated submanifold) that has codimension four.
This argument gives a good motivation for looking at
higher dimensional instanton on the ${\bf R}^4$ bundle.

Fortunately an eight dimensional metric that is prepared for this
purpose has been provided in \cite{PP}, \cite{GPP} and \cite{BFK}.
It is a metric of $Spin(7)$ holonomy
on the ${\bf R}^4$ bundle over $S^4$, which is an example of
a quaternionic line bundle over a quaternionic K\"ahler
manifold. In this paper we consider the octonionic Yang-Mills instanton on
this $Spin(7)$ holonomy manifold. As will be shown in the next section
this metric is a natural eight dimensional generalization of the Eguchi-Hanson
metric. The geometry of quaternions replaces the role of complex numbers
in the Eguchi-Hanson space whose global structure is the canonical line
bundle over the complex projective space $P_1({\bf C}) \cong S^2$.
Thus we can say that our instanton is a higher
dimensional generalization of anti-self-dual instanton on the ALE space.
It is known that the spinor bundle over
a four dimensional spin manifold has $Spin(7)$ holonomy in general.
The quaterninonic line bundle in this paper is identified
with the spinor bundle over $S^4$ (see Appendix A).

The paper is organized as follows;
in section two we introduce the metric of Gibbons-Page-Pope and set up
notations that are necessary in the following sections. The metric of
$Spin(7)$ holonomy is a solution to the octonionic self-duality for
spin connections. We show that the same condition is obtained as
a flow equation in a SUSY quantum mechanics of weight functions
appearing in our ansatz of the metric. In a course of explaining
the implication of the octonionic self-duality, some algebraic properties
of $Spin(7)$ as a subalgebra of $SO(8)$ are reviewed briefly.
We propose our ansatz for $Spin(7)$ Yang-Mills field in section three,
and write down the octonionic Yang-Mills instanton equation.
To solve the instanton equation we first make a classification of
solutions according to the asymptotic
behavior at infinity and find six classes.
There are reduced solutions in the sense that the actual gauge group is
reduced to
a direct product of $SU(2)$ factors in $Spin(7)$. For these class of
solutions we present almost complete answer in section four.
Unfortunately for other solutions we are not able to find
solutions in analytically closed form. In section five we perform
asympototic expansion
to see the existence of general solutions.
The final section is devoted to discussion.
We point out the relation to seven dimensional Chern-Simons theory. We also
make a remark on the energy-momentum tensor of higher dimensional Yang-Mills
instantons.

\section{Gravitational instanton in eight dimensions}
\renewcommand{\theequation}{2.\arabic{equation}}\setcounter{equation}{0}

We first derive a metric of $Spin(7)$ holonomy from the viewpoint of
supersymmetric quantum mechanics. This metric was originally obtained
by Gibbons-Page-Pope \cite{PP}\cite{GPP} and further discussed in \cite{BFK}.
(See also \cite{BS} for more intrinsic definition of the metric.)
We take the following ansatz for a metric on the ${\bf R}^4$ bundle over $S^4$
\beq
d\hat s^2 = f^2 dr^2 + g^2 ds^2 + h^2 ( \sigma_i - A_i)^2~, \label{metric}
\eeq
where
\beq
ds^2 = d\mu^2 + \frac{1}{4} sin^2 \mu \cdot \Sigma_i^2
\eeq
is the standard metric on the base space $S^4$. We assume that
$f, g$ and $h$ are functions of the radial coordinate $r$.
$\Sigma_i$ and $\sigma_i$ are left-invariant one-form of $SU(2)$ manifold;
\beq
d\Sigma_i = -\frac{1}{2} \e_{ijk} \Sigma_j \wedge \Sigma_k~, \quad
d\sigma_i = -\frac{1}{2} \e_{ijk} \sigma_j \wedge \sigma_k~.
\eeq
Note that $SU(2) \cong Sp(1)$ is the space of quaternions with unit norm.
Finally $A_i$ represents the basic $SU(2)$ instanton on $S^4$;
\beq
A_i = \cos^2 \frac{\mu}{2} \cdot \Sigma_i~.
\eeq
The vielbein (orthonormal frame) of the above metric is
\beqa
e^i = \frac{1}{2} g(r) \sin \mu \cdot \Sigma_i~, &\qquad&
e^{\hi} = h(r) \left( \sigma_i - A_i \right)~, \CR
e^7 = f(r) dr~,  &\qquad& e^8 = g(r) d\mu~,
\eeqa
where $i= 1,2,3$ and $\hi = 4,5,6= \hat 1 ,\hat 2, \hat 3$.
The indices $(1,2,3,8)$ are for the base space $S^4$ and
$(4,5,6,7)$ are those for the fibre.
This metric is a special case of more general class of
metric on the quaternionic line bundle over a quaternionic
K\"ahler manifold. Note that $S^4 = P_1({\bf H})$ is not a
K\"ahler but a quaternionic K\"ahler manifold. In fact
any four dimensional manifold is quaternionic K\"ahler.
The Eguchi-Hanson metric is a four dimensional gravitational
instanton on a (complex) line bundle over the complex
projective space $P_1({\bf C})$. More precisely the global
topology of the Eguchi-Hanson space is the cotangent bundle
$T^*(P_1({\bf C}))$ which coincides with the canonical bundle.
Hence, the above metric is
a higher dimensional generalization of the Eguchi-Hanson metric
obtained by simply replacing the complex numbers with
the quaternions (see also Appendix A).

It is straightforward to compute the Ricci tensor of the metric (\ref{metric}),
which is found to be diagonal;
\beqa
Ric_{ij} &=& \delta_{ij} \left\{ \frac{3}{g^2}
\left( 1 - \frac{1}{2} \frac{h^2}{g^2} \right) -4 K^2 - \frac{K'}{f}
-3KL \right\}~, \CR
Ric_{\hi\hj} &=& \delta_{ij} \left\{ \frac{h^2}{g^4} + \frac{1}{2h^2}
-\frac{L'}{f} -4KL -3 L^2 \right\}~, \label{Ricci} \\
Ric_{77} &=& -4 \frac{K'}{f} -4 K^2 - \frac{3}{f} L' -3 L^2~, \CR
Ric_{88} &=& Ric_{ii} \nonumber
\eeqa
where
\beq
K = \frac{g'}{fg}~, \qquad L = \frac{h'}{fh}~,
\eeq
and $'$ denotes the differentiation.
The volume form of the manifold is given by
\beq
e^1 \wedge e^2 \wedge \cdots \wedge e^8 = fg^4h^3 dr \wedge d\Omega_{S^7}~,
\eeq
where $d\Omega_{S^7}$ is the volume form of the sphere $S^7$.
We obtain the Einstein-Hilbert action;
\beqa
\int R \sqrt{g} d^8 x &=&vol_{S^7} \cdot (T + V)~, \CR
T &=& \int f^{-1} dr \left( 12 (g')^2 g^2 h^3 + 6 (h')^2 g^4 h + 24 (g'h')
g^3 h^2 \right) \\
V &=& \int f dr \left( 12 g^2 h^3 - \frac{3}{2} g^4 h -3 h^5 \right)~.
\eeqa
Note that we are going to regard the radial coordinate $r$ as a \lq\lq
time\rq\rq\
and hence the relative sign of the kinetic term and the potential term in
the action
is changed due to the Euclidean time. With this interpretation
the function $f(r)$ is a gauge freedom of time reparametrization.
In fact we would be able to impose a gauge fixing condition $f=1$.
Hence the physical variables are $g=g(r)$ and $h=h(r)$.
The (sigma model) metric determined from the kinetic term $T$ is
\beq
G_{gg} = 24 f^{-1}g^2 h^3~, \quad
G_{gh} = G_{hg} = 24 f^{-1}g^3 h^2~, \quad
G_{hh}= 12 f^{-1} g^4 h~.
\eeq
Now we are ready to make a crucial observation that
the present model can be identified as a bosonic part of
supersymmetric quantum mechanics ($0+1$ dimensional SUSY
sigma model).  The point is that introducing the superpotential
\beq
W = 3 g^4 h^2 + 6 g^2 h^4~,
\eeq
we can write the potential term in the form;
\beq
V = \frac{1}{2} G^{ij}~\frac{\del W}{\del q^i} \frac{\del W}{\del q^j}~,
\eeq
where $(q^1, q^2)=(g,h)$ and $G^{ij}$ is the inverse of the metric.
The equation of motion from the Hamiltonian\footnote{Note again that
the change of the relative sign in the Euclidean time.} $H = T - V$ of SUSY
quantum mechanics
\beq
\frac{d}{dr}q^i = - G^{ij} \frac{\del W}{\del q^j}
\eeq
gives the following flow equations;
\beq
K = \frac{g'}{fg} = - \frac{3}{2} \frac{h}{g^2}~, \quad
L = \frac{h'}{fh} = - \frac{1}{2h} + \frac{h}{g^2}~. \label{grvinst}
\eeq
If we introduced fermionic variables that are SUSY partners to  $(q^1,
q^2)=(g,h)$,
these equations would be equivalent to the condition that
SUSY variations of the fermionic coordinates vanish and hence
determine a purely bosonic configuration that is invariant under supersymmetry.
We observe that (\ref{grvinst}) coincides with the condition of the octonionic
self-duality of the Riemann curvature, which shows the BPS nature of
the octonionic self-duality. Using (\ref{Ricci}) and (\ref{grvinst}),
we can see that a solution to the above equation gives
a Ricci-flat metric. It is known that the same structure arises
for the four dimensional hyperK\"ahler metrics that depends only on
a radial coordinate regarded as an Euclidean time
in SUSY quantum mechanics. We note that what we have shown 
for the $Spin(7)$ holonomy is also valid for 
a seven dimensional metric with $G_2$ holonomy in \cite{GPP}.
It is an interesting problem to work out a similar relation to SUSY quantum
mechanics for other examples of the metric with special holonomy.

In \cite{BFK} the equations (\ref{grvinst})
are derived from the self-duality on the spin connection
\beq
\omega_{ab} = \frac{1}{2} \Psi_{abcd} \omega^{cd}~, \label{spininst}
\eeq
where a totally antisymmetric tensor $\Psi_{abcd}$ is defined
in Appendix B in terms of
the structure constants of octonions.
Imposed with a gauge condition on $g(r)$ instead of $f(r)$,
the equations (\ref{grvinst}) were solved as follows;
\beqa
f(r) &=& \left( 1- \left( \frac{m}{r} \right)^{10/3}\right)^{-1/2}~, \CR
g(r)^2 &=& \frac{9}{20} r^2~, \CR
h(r) &=& - \frac{3}{10} r f(r)^{-1}~. \label{BFK}
\eeqa
We note that $f(r)$ satisfies the equation
\beq
r f'(r) = \frac{5}{3} f(1 - f^2)~.
\eeq
In eight dimensions the spin connection $\omega$ is $SO(8)$ valued.
Let $\Gamma_{ab}$ be a generator of $SO(8)$.
The tensor $\Psi_{abcd}$ obeys the identity (\ref{octid}) which implies
\beq
\Psi_{abpq}\Psi^{cdpq} = 6 (\delta_a^c\delta_b^d
- \delta_a^d\delta_b^c) -4 \Psi_{ab}^{~~cd}~.
\eeq
This identity means that if we regard $\Psi_{ab}^{~~cd}$ as
a linear map $D$ on $SO(8)$ algebra, then the eigenvalues
of $(1/2)D$ are $1$ and $-3$. Since ${\hbox {dim}}~SO(8) = 28$ and
$D$ is traceless, we get the eigenspace decomposition;
\beq
SO(8) = E(1) \oplus E(-3)~,\quad ({\hbox {dim}}~E(1)= 21)~,
\eeq
where $E(1)$ coincides with $Spin(7)$ subgroup of $SO(8)$ \cite{DN}.
The orthogonal projection operator to each eigenspace is
\beq
P_1 = \frac{3}{4} \left( 1 + \frac{1}{6} D \right)~, \quad
P_{-3}= \frac{1}{4} \left( 1 - \frac{1}{2} D \right)~.
\eeq
We obtain the following generator of $Spin(7)$
\beq
G_{ab} = \frac{3}{4} \left( \Gamma_{ab} + \frac{1}{6} \Psi_{abcd}
\Gamma^{cd} \right)~,
\eeq
which satisfies the constraint
\beq
G_{ab} - \frac{1}{2} \Psi_{abcd} G^{cd} = 0~. \label{kousoku}
\eeq
The algebra $SO(8)$ has four mutually commuting $SU(2)$ subalgebras
with the following generators;
\beqa
S^{i}&=&(\Gamma^{8i}+\frac{1}{2}\e_{ijk}\Gamma^{jk}), \quad
T^{i}=(\Gamma^{8i}-\frac{1}{2}\e_{ijk}\Gamma^{jk}), \\
U^{i}&=&(\Gamma^{7\hi}+\frac{1}{2}\e_{ijk}\Gamma^{\hj\hk}), \quad
V^{i}=(\Gamma^{7\hi}-\frac{1}{2}\e_{ijk}\Gamma^{\hj\hk}).
\eeqa
By comparing the Dynkin diagrams of $SO(8)$ and $Spin(7)$, we see that
$Spin(7)$ is obtained by identifying two of $SU(2)$ subalgebras
which correspond to the outer nodes of $SO(8)$ diagram.
Indeed the constraint (\ref{kousoku}) implies an identification of
$S^i \equiv U^i$, that is $SU(2)$ in the base
direction and that along the fibre (the normal direction) are
identified\footnote{This is definetely related to the topological twist on
the world volume of SUSY cycles \cite{BSV}.}.
Since $\omega_{ab}\Gamma^{ab}=\omega_{ab}(P_1+ P_{-3})\Gamma^{ab}=
\omega_{ab}G^{ab}+ P_{-3}\omega_{ab}\Gamma_{ab}$,
we can regard an $SO(8)$ valued connection
with the octonionic self-duality $P_{-3}\omega_{ab} =0$
as $Spin(7)$ valued.  Thus the self-duality (\ref{spininst}) of the spin
connection imposed in \cite{BFK} is just the requirement
that $\omega$ is $Spin(7)$ valued. In fact we can prove that
the Cayley four form
\beq
\Omega = \frac{1}{4!} \Psi_{abcd} e^a \wedge e^b \wedge
e^c \wedge e^d~,
\eeq
is closed, if the spin connection satisfies the self-duality (\ref{spininst}).
The closedness of the Cayley four form is equivalent to
the condition of $Spin(7)$ holonomy \cite{Joy}\cite{SALA}. The total space of
the quaternionic line bundle on $S^4$ with the metric
given by (\ref{BFK}) is a manifold of $Spin(7)$ holonomy.

\section{Ansatz for $Spin(7)$ connection}
\renewcommand{\theequation}{3.\arabic{equation}}\setcounter{equation}{0}

Now we consider the octonionic instanton on a $Spin(7)$ bundle
over the GPP-BFK metric (octonionic gravitational instanton),
which we would like to propose as a higher dimensional generalization
of $SU(2)$ instanton on the Eguchi-Hanson metric (four dimensional
gravitational instanton.) In the four dimensional case,
the spin connection of the Eguchi-Hanson metric satisfies
the self-duality condition with respect to internal $SU(2)$ indices
and this implies the curvature two form satisfies
the self-duality on space-time indices as well as internal ones.
Thus we can obtain an example of Yang-Mills instanton by the standard
embedding of the spin connection into the Yang-Mills connection.
The same thing takes place for $Spin(7)$ case in eight dimensions
due to the basic identity for the structure constants of
octonions. (For $SU(2)$ case in the above the corresponding object
is $\e_{ijk}$; the structure constants of the quaternions.)
For the standard emdedding of $Spin(7)$ connections, see Appendix B.

We now try to find more general $Spin(7)$ instantons other than
the one obtained by the standard embedding.
Motivated by the form of spin connection of GPP-BFK metric,
we take the following ansatz
for $Spin(7)$ connection
\beqa
A_{ij} &=& - \left( \frac{1}{g \sin \mu} e^k + Y e^{\hk}
\right) \e_{ijk} \qquad
A_{\hi\hj} =  - \left( (X+Y) e^{\hk} +
\frac{\cot \frac{\mu}{2}}{g} e^k \right) \e_{ijk} \label{bansatz} \\
A_{i\hj} &=&  Z \left( \e_{ijk} e^k
+ \delta_{ij} e^8 \right)  \qquad
A_{8i} =  -\frac{\cot\mu}{g} e^i - Y e^{\hi} \\
A_{8\hi} &=& - Z e^i \qquad
A_{87} = - (3Z) e^8 \\
A_{7i} &=&  (3Z) e^i \qquad
A_{7\hi} = (X-Y) e^{\hi}~, \label{eansatz}
\eeqa
where $X(r), Y(r), Z(r)$ are unknown functions of the radial coordinate $r$.
For the spin connection we have
\beq
X(r) = \frac{1}{2h(r)} - \frac{h(r)}{2g(r)^2}, \quad
Y(r) = Z(r) = \frac{h(r)}{2g(r)^2}~.
\eeq
Since this ansatz satisfies the octonionic self-duality
\beq
A_{ab} = \frac{1}{2} \Psi_{abcd} A^{cd}~,
\eeq
with respect to the Lie algebra indices, we can consistently
regard the connection $A=\frac{1}{4}G^{ab}A_{ab}$ as $Spin(7)$ valued.
We also note that the vielbein of radial direction $e^7$ does
not appear, hence if we think of the radial coordinate
as a \lq\lq time\rq\rq, we are in the temporal gauge.

Due to the identity obeyed by the component of
the calibration four form or the structure constant of
octonions, the curvature $F= dA + A\wedge A$
satisfies the same self-duality as the connection and
hence it is also $Spin(7)$ valued. (That is, we can compute the curvature
as if $A$ was $SO(8)$ connection.)
We obtain the following curvature;
\beqa
F_{ij} &=& - \frac{1}{f} \left( Y' + \frac{h'}{h} Y \right)
e^7\wedge e^{\hk} \e_{ijk}
+ \left( \frac{1}{g^2} - \frac{h}{g^2}Y -10Z^2
\right) e^i \wedge e^j  \CR
& &~~ + \left( \frac{1}{h}Y -2Y^2
\right) e^{\hi} \wedge e^{\hj}
+ \left( -\frac{h}{g^2}Y +2Z^2
\right) e^8\wedge e^k \e_{ijk}  \label{bcurv} \\
F_{\hi\hj} &=&  - \frac{1}{f} \left( X' + Y' + \frac{h'}{h} (X + Y) \right)
e^7\wedge e^{\hk} \e_{ijk}
+ \left( \frac{1}{g^2} -
\frac{h}{g^2}(X+Y) -2Z^2 \right) e^i \wedge e^j  \CR
&+&\left( \frac{1}{h}(X+Y) -2X^2 -2Y^2
\right) e^{\hi} \wedge e^{\hj}
+ \left(  \frac{1}{g^2} - \frac{h}{g^2}(X+Y) -2Z^2
\right) e^8\wedge e^k \e_{ijk} \\
F_{i\hi} &=& \frac{1}{f} \left( Z' + \frac{g'}{g} Z\right)
e^7\wedge e^8
+ Z(4Y-3X)  e^i \wedge e^{\hi}
+  X Z \sum_{k\neq i} e^k \wedge e^{\hk} \\
F_{i\hj} &=& \frac{1}{f} \left( Z' + \frac{g'}{g} Z\right)
e^7\wedge e^k \e_{ijk}
+ Z(4Y-3X)  e^i \wedge e^{\hj}
- X Ze^j \wedge e^{\hi}  \CR
& &~~ -  X Z e^8\wedge e^{\hk} \e_{ijk} \qquad (i\neq j) \\
F_{8i} &=& - \frac{1}{f} \left( Y' + \frac{h'}{h} Y \right)
e^7\wedge e^{\hi} +
\left(  \frac{1}{g^2} - \frac{h}{g^2}Y
- 10 Z^2 \right) e^8 \wedge e^i  \CR
& &~~ + \left(- \frac{h}{2g^2}Y + Z^2 \right)
\e_{ijk} e^j \wedge e^k
+ \left( \frac{1}{2h}Y - Y^2 \right)
\e_{ijk} e^{\hj} \wedge e^{\hk} \\
F_{8\hi} &=& - \frac{1}{f} \left( Z'+ \frac{g'}{g} Z\right)
e^7\wedge e^i
+ Z (4Y -3X) e^8 \wedge e^{\hi}
- X Z \e_{ijk} e^j \wedge e^{\hk} \\
F_{87} &=& - \frac{3}{f} \left( Z'+ \frac{g'}{g} Z\right)
e^7\wedge e^8
+ Z (X -4Y) \sum_{k} e^k \wedge e^{\hk} \\
F_{7i} &=& \frac{3}{f} \left( Z'+ \frac{g'}{g} Z\right)
e^7\wedge e^i
+ Z (4Y -X) \e_{ijk} e^j \wedge e^{\hk}
- Z (4Y -X) e^8 \wedge e^{\hi} \\
F_{7\hi} &=& \frac{1}{f} \left( X' - Y' + \frac{h'}{h} (X-Y) \right)
e^7\wedge e^{\hi} + \left( \frac{h}{2g^2} (X-Y) - 3Z^2
\right) \e_{ijk} e^j \wedge e^k \CR
& &~~ +\left( -\frac{1}{2h} (X-Y) + X^2 - Y^2
\right) \e_{ijk} e^{\hj} \wedge e^{\hk}
 +  \left( \frac{h}{g^2} (X-Y) - 6Z^2
\right) e^8 \wedge e^i  \label{ecurv}
\eeqa

In the case of the Riemannian curvature which comes from the metric
the symmetry of four indices implies that the octonionic self-duality
as a space-time two form follows from that of $SO(8)$ indices.  However,
for the Yang-Mills curvature the octonionic instanton equation
\cite{CDFN}\cite{FN1}\cite{HK};
\beq
* F = \Omega \wedge F~,
\eeq
is independent of the self-duality of $F$ with respect to
the Lie algebra indices which only implies that it is $Spin(7)$ valued.
Thanks to the cyclic symmetry of $SU(2)$ indices in our ansatz,
the octonionic self-dualty of the curvature two-from is greatly reduced.
We find the following three independent conditions;
\beqa
\frac{1}{f} X' + L \cdot X +2X^2 - \left( \frac{1}{h} + \frac{2h}{g^2}\right) X
+ \frac{1}{g^2} + 4 Z^2 &=& 0~, \CR
\frac{1}{f} Y' + L \cdot Y +2Y^2 - \left( \frac{1}{h} + \frac{2h}{g^2}\right) Y
+ \frac{1}{g^2} - 8 Z^2 &= & 0~,\\
\frac{1}{f} Z' + K\cdot Z + (X-4Y) Z &=& 0~. \nonumber
\eeqa

\section{Reduced Solutions}
\renewcommand{\theequation}{4.\arabic{equation}}\setcounter{equation}{0}

Using the explicit form of the functions $g$ and $h$ in (\ref{BFK})
and making the following change of variables which is convenient
for this special background of gravitational instanton
\beqa
X &=& \frac{1}{rf}\left( \xi_X-\frac{5}{3}f^2 \right), \label{henkanx}\\
Y &=& \frac{1}{rf}\left( \xi_Y-\frac{5}{3}f^2 \right), \label{henkany}\\
Z &=& \frac{1}{rf}\xi_Z, \label{henkanz}
\eeqa
we obtain the following system of first order ordinary differential equations :
\beqa
r\frac{d}{dr}\xi_{X}+2\xi_{X}(\xi_X-1)
+4\xi_{Z}^2&=&0~, \label{boctinst} \\
r\frac{d}{dr}\xi_{Y}+2\xi_{Y}(\xi_Y-1)
-8\xi_{Z}^2&=&0~, \label{bboctinst} \\
r\frac{d}{dr}\xi_{Z}+(\xi_X-4\xi_Y+5)\xi_{Z}
+\frac{20}{3}(f^2-1)\xi_{Z}&=&0~. \label{eoctinst}
\eeqa

Before considering solutions to the differential equations
(\ref{boctinst})-(\ref{eoctinst})
let us first look at the asymptotic behavior at infinity.
In the region of the large radial coordinate $r$
neglecting the last term in (\ref{eoctinst}) (or putting $f=1$), we obtain
gradient flow equations on ${\bf R}^3$=$\bigl\{\xi=(\xi_X,\xi_Y,\xi_Z) \bigr\}$
with a metric $\eta_{AB}=diag(1/2,1,4)$:
\beq
\frac{d}{dt}\xi^A=- \eta^{AB} \frac{\partial U(\xi)}{\partial \xi^B}~,
\label{gradflow}
\eeq
where $t=\ln r$ and the potential function $U$ is given by
\beq
U=\frac{1}{3}\xi_{X}^3-\frac{1}{2}\xi_{X}^2
+\frac{2}{3}\xi_{Y}^3-\xi_{Y}^2
+2\xi_{Z}^2(\xi_X-4\xi_Y+5).
\eeq
\begin{table}
\caption{Critical points of the potential function $U$ and their Morse indices}
\begin{center}
\begin{tabular}{|c|c|c|}
\hline
critical point & critical values & index \\
\hline
$P_1$ & (1,1,0) & 0 \\
\hline
$P_2$ & (1,0,0) & 1 \\
\hline
$P_3$ & (0,1,0) & 1 \\
\hline
$P_4$ & (0,0,0) & 2 \\
\hline
$P_5^\pm$ & $(1/3,4/3,\pm 1/3)$ & 2 \\
\hline
$P_6^\pm$ & $(5/11,15/11,\pm \sqrt{15}/11)$ & 1 \\
\hline
\end{tabular}
\end{center}
\end{table}
By counting the number of negative eigenvalues of the Hessian
\beq
H_{AB} = \frac{\partial^2 U}{\partial\xi_A \partial\xi_B}
= \left( \begin{array}{ccc}
2\xi_X -1 &  0 &  4\xi_Z \\
 0 &  4\xi_Y -2  & -16\xi_Z  \\
4\xi_Z  &  -16\xi_Z  &  4(\xi_X -4\xi_Y +5) \\
\end{array} \right)
\eeq
at the critical points of $U$,
we find the list of the critical points and the Morse indices (see Table 1).
We can use one of these critical points as a boundary condition of the
octonionic Yang-Mills instantons and classify
the solutions of (\ref{boctinst})-(\ref{eoctinst}) according to
\beq
Sol(P_i) := {\hbox{\{a family of solutions approaching to $P_i$
for $r \rightarrow \infty$ \}. }}
\eeq

If $\xi_Z=0$, we can neglect the last term
of (\ref{eoctinst}) in the whole region and
we can find a general solution:
\beq
\xi_X=\frac{1}{1-\frac{a}{r^2}},\quad
\xi_Y=\frac{1}{1-\frac{b}{r^2}}, \label{zosol}
\eeq
which belongs to the class $Sol(P_1)$ for finite parameters
$(a,b)$. When we take the limits of
parameters $(a,\infty), (\infty,b)$ and $(\infty,\infty)$, the limitting
solutions
belong to $Sol(P_2), Sol(P_3)$ and $Sol(P_4)$, respectively.
When $\xi_Z \neq 0$ it is difficult to find general solutions, which will be
discussed in more detail in the next section, but we here present a
special solution in $Sol(P_{5}^{-})$ corresponding to the spin connection
\beq
\xi_X=\frac{1}{3}, \quad
\xi_Y=-\frac{1}{3}+\frac{5}{3}f^{2}, \quad \xi_Z=-\frac{1}{3}. \label{standard}
\eeq
Note that provided a solution $(\xi_X,\xi_Y,\xi_Z)\in Sol(P_{5,6}^{+})$
there always exists a solution $(\xi_X,\xi_Y,-\xi_Z)\in Sol(P_{5,6}^{-})$
by the symmetry of equations.

The spin connection (\ref{standard}) yields a $Spin(7)$ connection,
while as we will see shortly the connection of (\ref{zosol})
takes the value in a sub-Lie algebra of $Spin(7)$.
In this sense (\ref{zosol}) is a reduced solution.
Let us calculate the curvature two-form
$F=\frac{1}{4}G^{ab}F_{ab}$ by substituting
 (\ref{henkanx}),(\ref{henkany}) and $Z=0$ into (\ref{bcurv})-(\ref{ecurv}).
Defining the sub-generators of $G^{ab}$ by
\beqa
S^{i}&=&\frac{1}{4}(-G^{7\hi}+\frac{1}{2}\e_{ijk}G^{\hj\hk}),
\\
T^{i}&=&\frac{1}{2}(G^{8i}+\frac{1}{2}\e_{ijk}G^{jk}) =
\frac{1}{2}(G^{7\hi}+\frac{1}{2}\e_{ijk}G^{\hj\hk}), \\
U^{i}&=&\frac{1}{4}(-G^{8i}+\frac{1}{2}\e_{ijk}G^{jk}),
\eeqa
we can rewrite the curvature into the following form:
\beq
F=S^{i}F_{i}(\xi_{X})+T^{i}F_{i}(\xi_{Y})+U^{i}C_{i},
\eeq
where the two-forms $F_{i}(\xi_A)$ $(A=X,Y)$ and $C_i$ are given by
\beq
F_{i}(\xi_A)=f_{A}\left( e^{8}\wedge e^i+
\frac{1}{2}\e_{ijk}e^{j}\wedge e^{k} \right)
+\frac{g_A}{2}
\e_{ijk}e^{\hj}\wedge e^{\hk}
+h_{A}e^7\wedge e^{\hi}, \label{exa}
\eeq
with
\beq
f_{A}=\frac{4\xi_A}{3r^{2}f^{2}}, \quad
g_{A}=-\frac{4\xi_A}{r^{2}f^{2}}\left( \xi_A-
\frac{5}{3}f^2 \right), \quad
h_{A}=\frac{4\xi_A}{r^{2}f^{2}}\left( \xi_A+\frac{2}{3}-
\frac{5}{3}f^2 \right),
\eeq
and
\beq
C_{i}=\frac{20}{9r^2}\left( -e^8\wedge e^{i}+\frac{1}{2}\e_{ijk}
e^j\wedge e^k \right). \label{exc}
\eeq
It is easy to see that $S^i, T^i$ and $U^i$ are mutually commuting
$SU(2)$ generators which satisfy the relations
\beq
[S^i, S^j]=-\e_{ijk}S^k, \;[T^i, T^j]=-\e_{ijk}T^k, \;
[U^i, U^j]=-\e_{ijk}U^k.
\eeq
Thus the solution (\ref{zosol}) describes the $SU(2)^3$ octonionic Yang-Mills
instanton.
When the parameters $(a,b)$ take the special values, some curvature
components vanish:
\beqa
F_{i}(\xi_Y)=0 \quad \mbox{for} \;(a,\infty), \label{reducea}\\
F_{i}(\xi_X)=0 \quad \mbox{for} \;(\infty,b), \label{reduceb}\\
F_{i}(\xi_X)=F_{i}(\xi_Y)=0 \quad \mbox{for} \label{reducec}\;(\infty,\infty).
\eeqa
The gauge group is then further reduced to $SU(2)^2$ and $SU(2)$ in the case of
(\ref{reducea})
or (\ref{reduceb}) and (\ref{reducec}), respectively.

We now evaluate the Chern forms characterizing the topological nature
of the solution (\ref{zosol}). The relevant closed forms are given by
\beqa
c_2=\frac{1}{8\pi^2}\, \trace~F\wedge F,\qquad \qquad \qquad \qquad \\
c_4=\frac{1}{128\pi^4}\,((\trace~F\wedge F)(\trace~F\wedge F)
-2\,\trace~F\wedge F\wedge
F\wedge F),
\eeqa
where $\trace$ refers to the adjoint representation of $SU(2)^3$.
Since $c_2$ is a four-form, it must be integrated over a four-dimensional
hypersurface in the quaternionic line bundle. A natural choice is the
fibre ${\bf R}^4$, which is specified by the orthonormal frame
$\{ e^4,e^5,e^6,e^7 \}$.
Using (\ref{zosol}), (\ref{exa}) and (\ref{exc}), we obtain the formula
\beq
\int \, c_2=\left( \frac{3}{10} \right)^{3}
\frac{3!}{\pi^2}vol_{SU(2)}
\sum_{A=X,Y} I_{A},
\eeq
where
\beqa
I_{A} &=& -\frac{1}{4} \int_{m}^{\infty} dr\,r^3 f^{-2}g_{A}h_{A} \nonumber \\
      &=& \frac{1}{3} \int_{m}^{\infty} dr\frac{d}{dr} \left(
5\xi_{A}^2 f^{-4}
-2\xi_{A}^3 f^{-6} \right)
\eeqa
in the parameter region $a,b\,<\,m^2$. Since
$f^{-1}=0$ at $r=m$ and $\xi_{X}=\xi_{Y}=f=1$ for $r \rightarrow \infty$,
we have $I_{X}=I_{Y}=1$. For $c_4$ the integration is to be evaluated over
the total space of the quaternionic line bundle. A similar calculation
yields
\beq
\int \, c_4 = \left( \frac{3}{10} \right)^5
\frac{4}{\pi^4}vol_{S^7}
(25 \sum_{A=X,Y} I_{A}-I_{XY}),
\eeq
where
\beqa
I_{XY} &=& -\left( \frac{3}{4} \right)^3 \sum_{A \neq B}
\int_{m}^{\infty} dr\,r^{7}f^{-2}
(3f_{A}^{2}g_{B}h_{B}+2f_{A}f_{B}g_{A}g_{B}) \nonumber \\
       &=& \int_{m}^{\infty} dr\,\frac{d}{dr}
(25 f^{-8}\xi_{X}^2 \xi_{Y}^2 - 6 f^{-10}\xi_{X}^2 \xi_{Y}^2
(\xi_{X}+\xi_{Y}))
\eeqa
and $I_{XY}=13$ for $a,b<m^2$.

\section{Asymptotic Expansion}
\renewcommand{\theequation}{5.\arabic{equation}}\setcounter{equation}{0}

As we have mentioned it is difficult to obtain a solution with $\xi_Z \neq 0$
in a closed form. The only exceptional solution is given by
the spin connection regarded as a $Spin(7)$ gauge field.
To investigate the existence of more general solutions let us look for a
solution in a form of formal power series.
We first consider the exponent of the power series
determined by the gradient flow equation.
Linearizing the equation (\ref{gradflow}) around the critical point $P_i$,
we have
\beq
\frac{d}{dt}\widetilde{\xi}_{A}= - \widetilde{H}_{AB} \widetilde{\xi}_{B},
\label{linear}
\eeq
where $\widetilde{H}_{AB}$ represents the $3 \times 3$ matrix $\eta^{AC}H_{CB}$
evaluated at $P_i$.
To be specific, let $\lambda_{1},\lambda_{2}, \cdots$
be the positive eigenvalues of $\widetilde{H}$ and
$\xi_{A}(P_i)$ the critical values.
Then, the gradient flow approaching to $P_i$ may be
expanded in the form
\beq
\xi_A=\xi_{A}(P_i)
+a_{1}^{A} \left( \frac{a}{r} \right)^{\lambda_{1}}
+a_{2}^{A} \left( \frac{b}{r} \right)^{\lambda_{2}}
+\cdots
\label{appro}
\eeq
for a large radial coordinate. Here coefficients
$a_{1}^{A},a_{2}^{A}, \cdot \cdot \cdot $
are determined by the linearized
gradient flow equation (\ref{linear})
and $a,b, \cdots$ are
$3-index(P_i)$ free parameters .
Recall that the gradient flow gives an
approximate solution, and hence the expansion (\ref{appro}) must be
corrected by the exact octonionic Yang-Mills equation.
The correction comes from the last term $(f^2-1) \xi_Z$ in (\ref{eoctinst}),
which is also expanded by using
\beq
f^2=\sum_{n=0}^{\infty} \left( \frac{m}{r} \right)^{10n/3}.
\eeq
Now we make an ansatz of the formal power series which takes into acount
of the asymptotic behavior described above:
\beq
\xi_A \approx \xi_{A}(P_i)+ \sum_{k=0}^{\infty}\sum_{n_{1}=0}^{\infty}
\sum_{n_{2}=0}^{\infty} \cdot \cdot \cdot (w^{k}z_{1}^{n_1}z_{2}^{n_2}
\cdot \cdot \cdot)\, a_{kn_{1}n_{2}\cdot \cdot \cdot }^{A},\quad
k+n_{1}+n_{2}+ \cdot \cdot \cdot \neq 0, \label{expan}
\eeq
where $w=\left( \frac{m}{r} \right)^{10/3}$, $z_{1}=\left( \frac{a}{r}
\right)^{\lambda_{1}}$, $z_{2}=\left( \frac{b}{r} \right)^{\lambda_{2}}$
$\cdot \cdot \cdot $.
This series will provide the asymptotic expansion of the octonionic
Yang-Mills instanton in $Sol(P_{i})$ with totally $4-index(P_i)$
moduli parameters, if the parameter $m$ of the background metric is
treated as a moduli. In the following we illustrate this asymptotic expansion
by concentrating on the cases $Sol(P_{5}^{-})$ and $Sol(P_1)$.
Other cases can be analyzed in the same manner.

\begin{flushleft}
(a) $Sol(P_{5}^{-})$
\end{flushleft}

The matrix $\widetilde{H}$ has one positive eigenvalue $\lambda =
\frac{1}{3}(7+\sqrt{57})$, so that the solution is written as the double
power series:
\beq
\xi_A \approx \xi_{A}(P_{5}^{-})+ \sum_{n=0}^{\infty}\sum_{k=0}^{\infty}
\, w^{n}z^{k} a_{nk}^{A},\quad
n+k \neq 0,
\eeq
where $w=\left( \frac{m}{r} \right)^{10/3}$, $z=\left( \frac{a}{r}
\right)^\lambda$ and we have one free  (moduli) parameter $a$ in addition to
the background metric moduli $m$.
When we take $a_{01}^X=a_{01}^Y=a_{01}^Z=0$ as an initial condition of
the recursion relation, we find a solution
\beq
a_{n0}^Y = \frac{5}{3}~~(\forall n \geq 1)~, \qquad {\hbox{others}}=0~,
\eeq
which recovers the spin connection (\ref{standard}).
More general solution with the additional moduli pamameter $a$ is
obtained by using the series
\beqa
\xi_X &\approx& \frac{1}{3}+ \sum_{n=0}^{\infty}\sum_{k=1}^{\infty}
\, w^{n}z^{k} a_{nk}, \\
\xi_Y &\approx& -\frac{1}{3}+\frac{5}{3}f^2+
\sum_{n=0}^{\infty}\sum_{k=1}^{\infty}
\, w^{n}z^{k} b_{nk}, \\
\xi_Z &\approx& -\frac{1}{3}+ \sum_{n=0}^{\infty}\sum_{k=1}^{\infty}
\, w^{n}z^{k} c_{nk},
\eeqa
where the first coefficients $a_{01}=-\frac{1}{3}(9-\sqrt{57})$,
$b_{01}=\frac{1}{3}(3+\sqrt{57})$ and $c_{01}=1$.
The higher coefficients are uniquely determined by the recursion formulae:
\beqa
\left( \frac{10}{3}n+\lambda k+\frac{2}{3} \right)a_{nk}+\frac{8}{3}
c_{nk}-2 \sum_{p=0}^{n}\sum_{q=1}^{k-1} (a_{pq}a_{n-p,k-q}+
2c_{pq}c_{n-p,k-q})=0, \\
\left( \frac{10}{3}n+\lambda k-\frac{10}{3} \right) b_{nk}
-\frac{16}{3}c_{nk}
-\frac{20}{3} \sum_{p=0}^{n-1} \,b_{pk} \qquad \qquad \qquad \qquad
\qquad \qquad \nonumber \\
-2 \sum_{p=0}^{n}\sum_{q=1}^{k-1} (b_{pq}b_{n-p,k-q}-
4c_{pq}c_{n-p,k-q})=0, \qquad \qquad \qquad \\
\frac{1}{3}a_{nk}-\frac{4}{3}b_{nk}+
\left( \frac{10}{3} n +\lambda k \right)c_{nk}-
\sum_{p=0}^{n}\sum_{q=1}^{k-1}\,c_{pq} (a_{n-p,k-q}-
4b_{n-p,k-q})=0.
\eeqa
We leave the issue of convergence of this formal solution for
future research. This is important for a global structure of
the moduli space such as a compactification.

\begin{flushleft}
(b) $Sol(P_{1})$
\end{flushleft}

The matrix $\widetilde{H}$ has degenerate eigenvalues, i.e., $\lambda_{1} =
\lambda_{2} = \lambda_{3} = 2$.
In order to unambiguously determine the coefficients
$a_{kn_{1}n_{2}n_{3}}^{A}$ of power series, we first perturb the octonionic
Yang-Mills equation so that the corresponding $\widetilde{H}$ has
non-degenerate eigenvalues and after all calculations we take off the
perturbation. If we use the metric
$\eta_{AB}=diag(\frac{1}{2}(1-\epsilon_{1}),1-\epsilon_{2},4)$
with small parameters $\epsilon_{1}$ and $\epsilon_{2}$ instead of
$\eta_{AB}=diag(\frac{1}{2},1,4)$,
then the positive eigenvalues of $\widetilde{H}$ are $\lambda_{1}=
2(1+\epsilon_{1}), \lambda_{2}=2(1+\epsilon_{2}), \lambda_{3}=2$.
The equations (\ref{boctinst}) and (\ref{bboctinst}) are replaced by
\beqa
r\frac{d}{dr}\xi_{X}+\lambda_{1}\xi_{X}(\xi_X-1)
+2\lambda_{1}\xi_{Z}^2&=&0~, \\
r\frac{d}{dr}\xi_{Y}+\lambda_{2}\xi_{Y}(\xi_Y-1)
-4\lambda_{2}\xi_{Z}^2&=&0~,
\eeqa
while the equation (\ref{eoctinst}) for the Z-component remains unchanged.
After taking the limit $\epsilon_{i} \rightarrow 0$ (i=1,2), the solution
in a form of power series is given by (\ref{expan}) with
$w=\left( \frac{m}{r} \right)^{10/3}$, $z_{1}=\left( \frac{a}{r} \right)^{2}$,
$z_{2}=\left( \frac{b}{r} \right)^{2}$ and
$z_{3}=\left( \frac{c}{r} \right)^{2}$.
The explicit lower terms are as follows:
\beqa
\xi_X &\approx& \frac{1}{1-z_{1}}+ \sum_{n=1}^{\infty}\,z_{3}^{n}a_{n}
+\cdot \cdot \cdot \\
\xi_Y &\approx& \frac{1}{1-z_{2}}+ \sum_{n=1}^{\infty}\,z_{3}^{n}b_{n}
+\cdot \cdot \cdot \\
\xi_Z &\approx& \sum_{n=1}^{\infty}\,z_{3}^{n}c_{n}
+\cdot \cdot \cdot
\eeqa
where the coefficients are determined by the recursion formulae
\beqa
a_{n} &=& \frac{1}{n-1} \sum_{k=1}^{n-1} (a_{k}a_{n-k}+
2c_{k}c_{n-k}), \\
b_{n} &=& \frac{1}{n-1} \sum_{k=1}^{n-1} (a_{k}a_{n-k}-
4c_{k}c_{n-k}), \\
c_{n} &=& \frac{1}{2(n-1)} \sum_{k=1}^{n-1} c_{k}(a_{n-k}-
4b_{n-k})
\eeqa
with the first terms $a_{1}=b_{1}=0$ and $c_{1}=1$.
If we impose $c=0$, the above expansion reproduces the exact solution
(\ref{zosol}).

\section{Discussion}
\renewcommand{\theequation}{6.\arabic{equation}}\setcounter{equation}{0}

Firstly we remark the relationship betweeen the asymptotic flow
(\ref{gradflow}) and the Chern-Simons theory over the squashed seven sphere
$\hat{S}^7$. A similar relation in various dimensions has been discussed in
\cite{BLN}.
The gravitational instanton metric (\ref{metric}) has the conformally product
form in a large radial coordinate region,
\beq
ds^2 \longrightarrow dr^2+\frac{9}{20}r^2 ds_{\hat{S}^7}^{2},
\eeq
where $ds_{\hat{S}^7}^{2}$ is the $Sp(2)\cdot Sp(1)$-invariant metric on
$\hat{S}^7$:
\beq
ds_{\hat{S}^7}^{2}=d\mu^{2}+\frac{1}{4}\sin^{2}\mu\cdot\Sigma_{i}^{2}
+\frac{1}{5}(\sigma_i-A_i)^{2}. \label{sphere}
\eeq
Now the equation (\ref{gradflow}) may be regarded as the gradient flow of
the Chern-Simons
theory on $\hat{S}^7$ as follows. Let us define the action by
\beq
CS[A]=\int_{\hat{S}^7}\hat{\Psi}\wedge
\trace~(A\wedge dA+\frac{2}{3}A \wedge A \wedge A),
\label{CS}
\eeq
where $A$ stands for $Spin(7)$ connection and $\hat{\Psi}$
a closed four-form on
$\hat{S}^7$ induced by the calibration four-form on the total space with
$Spin(7)$ holonomy. It is explicitly written as
\beqa
\hat{\Psi}&=&\theta^{1}\wedge \theta^{2}\wedge \theta^{3}\wedge \theta^{8}
-\theta^{1}\wedge \theta^{\hat{2}}\wedge
\theta^{\hat{3}}\wedge \theta^{8}
+\theta^{2}\wedge \theta^{\hat{1}}\wedge \theta^{\hat{3}}\wedge \theta^{8}
-\theta^{3}\wedge \theta^{\hat{1}}\wedge
\theta^{\hat{2}}\wedge \theta^{8} \nonumber \\
& &~~~+\theta^{1}\wedge \theta^{2}\wedge
 \theta^{\hat{1}}\wedge \theta^{\hat{2}}
+\theta^{2}\wedge \theta^{3}\wedge \theta^{\hat{2}}\wedge \theta^{\hat{3}}
+\theta^{1}\wedge \theta^{3}\wedge
\theta^{\hat{1}}\wedge \theta^{\hat{3}},
\eeqa
using the orthonormal frame of the metric (\ref{sphere})
\beq
\theta^{i}=\frac{1}{2}\sin\mu\cdot\Sigma_{i},\quad
\theta^{\hi}=\frac{1}{\sqrt{5}}(\sigma_{i}-A_{i}),\quad \theta^{8}=d\mu.
\eeq
The gradient flow of the functional (\ref{CS}) obeys a differential equation
\beq
\frac{d}{dt}A=-*(\hat{\Psi}\wedge F),
\eeq
which reproduces the flow (\ref{gradflow}) if we use
the Ansatz (\ref{bansatz})-(\ref{eansatz}) and (\ref{henkanx})-
(\ref{henkanz}) with $f=1$. The critical points
of $CS[A]$, i.e., the solutions of $\hat{\Psi}\wedge F=0$, then reduce to
those of the potential function $U$,
which does not mean flat-connections as in the 3-dimensional
Chern-Simons theory.

The energy-momentum tensor of the Yang-Mills field is
\beq
T^{YM}_{\mu\nu} = \trace (F_{\mu\lambda}F_\nu^{~\lambda}) -
\frac{1}{4}g_{\mu\nu}
\trace (F_{\lambda\kappa}F^{\lambda\kappa})~. \label{em}
\eeq
For the octonionic instantons the energy-momentum tensor $T_{\mu\nu}$ does
not vanish in general and this presents a sharp contrast with the case of
four dimensional instantons. We can prove that (anti) self-duality of the
curvature
implies $T^{YM}=0$ using a property of the totally antisymmetric tensor
$\e_{\mu\nu
\rho\sigma}$ in four dimensions. Hence a four dimensional instanton does not
disturb the Ricci flatness of the background metric. However the octonionic
self-duality is not sufficient for leading a vanishing energy-momentum tensor
and this causes an issue of the back reaction of matter to gravity.
Non-vanishing total energy-momentum tensor is inconsistent with
the Ricci flatness of the space-time. Actually a similar issue is encountered,
if we embed a four dimensional instanton in higher dimensions. (Note that
this gives a special case of higher dimensional instanton.) In such a case,
though the tangent components of $T_{\mu\nu}$ along the four dimensional
submanifold on which the instanton lives vanish as we argued above,
the instanton produces a non-vanishing contribution to the $(d-4)$ dimensional
normal components of the energy-momentum tensor.
One of the ways to resolve this problem is to embed the solution to
a consistent background of superstrings. Based on a four dimensional instanton,
one can construct several five brane solutions which has a dilaton $\phi$
and an anti-symmetric tensor field $H = dB$ in addition to the Yang-Mills
field $A$ \cite{Sto}. In these solutions we can think of four dimensional
instanton as a source of configuration in the transverse space.
Due to a coupling
\beq
dH = \trace ( R \wedge R - F \wedge F)
\eeq
there are contributions from $\phi$ and $B$ which balances the
total energy-momentum tensor. This should be so,
since this provides a consistent background for superstrings.
>From this example we believe our solutions should be promoted to
a consistent background of supermembrane or eleven dimensional
super gravity \cite{HS}\cite{DEKLM}\cite{BFK}.

\vskip10mm

\begin{center}
{\bf Acknowledgements}
\end{center}

We would like to thank Y.Hashimoto, T.Ootsuka and S.Miyagi for useful
discussions.
The work of H.K. is supported in part by the Grant-in-Aid
for Scientific Research on Priority Area 707
"Supersymmetry and Unified Theory of Elementary Particles" and
 No. 10640081, from Japan Ministry of Education.

\newpage

\noindent

{\Large\bf Appendix A. Quaternionic K\"ahler Manifold}

\renewcommand{\theequation}{A.\arabic{equation}}\setcounter{equation}{0}

\vskip3mm\noindent
The holonomy group of a connected and orientable
$4n$ dimensional Riemannian manifold is a subgroup of $SO(4n)$.
If the holonomy group is reduced to $Sp(n)\cdot Sp(1) \cong Sp(n)
\times Sp(1)/ {\bf Z}_2$,
the manifold is called quaternionic K\"ahler \cite{SALA}.
(Caution: a quaternionic K\"ahler manifold is not necessarily
K\"ahler !)
The quaternionic K\"ahler manifold is known as
a target space geometry of $N=2$ supergravity in four dimensions.
(If we consider $N=2$ global SUSY, then $Sp(1)$ part is trivial and
the manifold is hyperK\"ahler.) It also provides a natural arena
for a higher dimensional generalization of instanton equation.
Note that when $n=1$ the holonomy of
quaternionic K\"ahler manifold is $Sp(1) \times Sp(1)/ {\bf Z}_2$
and hence any four dimensional orientable manifold is quaternionic K\"ahler.
This may be compared with the fact any two dimensional
orientable manifold is K\"ahler due to $SO(2) \cong U(1)$.

On the frame bundle of a quaternionic K\"ahler manifold
the gauge transformation among local coordinate patches is
in $Sp(n)\cdot Sp(1)$. We can define a quaternionic line bundle
as an associated bundle to the frame bundle as follows;
Each fibre of quaternionic line bundle is the space of
quaternions $\bf H$ that is ${\bf R}^4$ as a vector space.
On quaternions there is a natural action of $Sp(1) \cong SU(2)$.
Thus we can define the gauge transformation of the fibre using
the $Sp(1)$ part of the holonomy group. That is $Sp(n)$ part
acts trivially on the quaternionic line bundle. This especially
implies that quaternionic line bundle on hyperK\"ahler manifold
is trivial. It might be helpful to note that this construction is
a quaternionic analogue of complex line bundle (or the canonical $U(1)$ bundle)
on a K\"ahler manifold that has $U(n)$ holonomy.

If we take a four dimensional spin manifold $M$,
each factor of $Sp(1)$ is identified as $Spin(3)$ that
defines the spinor bundle $S^\pm(M)$. Hence, the complexified
spinor bundle on four dimensional manifold is regarded as a quaternionic
line bundle. It is known that the total space of spinor bundle
on a four dimensional manifold possesses a natural $Spin(7)$ structure.

\vskip10mm

{\Large\bf Appendix B. Standard Embedding of $Spin(7)$ connection}

\renewcommand{\theequation}{B.\arabic{equation}}\setcounter{equation}{0}

\vskip3mm\noindent
We show that the octonionic self-duality of the spin connection $\omega$
with respect
to the (local frame) Lie algebra indices implies the octonionic self-duality of
the curvature $R = d\omega + \omega \wedge \omega$ with respect to the
space time
form indices. Let $\omega_{ab} = - \omega_{ba}$ be the spin connection one-form.
In eight dimensions $\omega$ is $SO(8)$ valued in general. (We take the
Euclidean
signature in the following.) But if we impose the octonionic self-duality
\beq
\omega_{ab} = \frac{1}{2} \Psi_{abcd} \omega^{cd}~,
\eeq
the spin connection $\omega$ can be regarded as $Spin(7)$ valued.
The totally anti-symmetric tensor $\Psi_{abcd}$ is related to
the structure constants $c_{abc}$ of the octonion algebra as follows \cite{DGT};
\beq
\Psi_{abc8}= c_{abc}~, \quad \Psi_{abcd} = \frac{1}{3!} \e_{abcdpqr}c^{pqr}~,
\eeq
where the duality is taken in seven dimensions.
In terms of a local frame (vielbein) $e_\mu^a$ we can define a space-time
self-dual four form
\beq
\Omega = \frac{1}{4!} \Psi_{abcd} e^a \wedge e^b \wedge e^c \wedge e^d~.
\eeq
On an eight dimensional manifold of $Spin(7)$ holonomy,
the four form $\Psi$ is closed;
\beq
d \Omega = 0~,
\eeq
A crutial point is the following
identity which follows from the property of the octonionic structure
constants $c_{abc}$ \cite{DGT}\cite{DN},
\beqa
\Psi_{abcd} \Psi^{fghd} &=&  ( \delta_a^f \delta_b^g - \delta_b^f
\delta_a^g) \delta_c^h
+ ( (fgh) : cyclic)   \label{octid} \\
& &  - \left( (\Psi_{ab}^{fg} \delta_c^h
 + \Psi_{bc}^{fg} \delta_a^h  + \Psi_{ca}^{fg} \delta_b^h)
 +  ( (fgh) : cyclic) \right)~.  \nonumber
\eeqa
Due to the symmetry of the Riemann curvature tensor
the self-duality of form indices follows form that of Lie algebra indices.
Thus it is enough to show that
\beq
R_{ab} = \frac{1}{2} \Psi_{abcd} R^{cd}~,
\eeq
where we have suppressed the two form indices.
In the defining relation
\beq
R_{ab} = d \omega_{ab} + \omega_{ac} \wedge \omega_{cb}
\eeq
the self-duality of the second term is non-trivial.
Using the identity (\ref{octid}), we can show the second term is
indeed self dual. Thus the curvature two form constructed from a $Spin(7)$
valued
spin connction $\omega$ satisfies the octonionic instanton equation.



\end{document}